\documentstyle[prd,aps,floats,epsf]{revtex}
\begin{document}
\title{ 
Adiabatic Gravitational Perturbation During Reheating
}
\author
{W.B. Lin$^{1}$, X.H. Meng$^{2}$ and Xinmin Zhang$^{3,1}$}
\address{~\\$^1$
Institute of High Energy Physics, Chinese 
Academy of Science, PO Box 918-4, Beijing 100039, P. R. China;
~\\$^2$
Department of Physics, Nankai University, Tianjin, 300071, P. R. China;
~\\$^3$
CCAST (World Lab), PO Box 8370, Beijing 100080, P. R. China}

\maketitle

\begin{abstract}
{
We study the possibilities of parametric amplification of
the gravitational perturbation during reheating in 
single-field inflation models. Our result shows that
there is no additional growth of the super-horizon 
modes beyond the usual predictions.
}
\end{abstract}


\vskip 0.4cm

\section{Introduction}

As was realized in [\ref{TB1}] that 
parametric resonance instability occurs during reheating period when the 
inflaton field $\phi$ oscillates. 
Since gravitational perturbation is coupled to the inflaton 
by the Einstein equation,
it may also experience parametric resonance
amplification during this stage. This issue has been studied in
Refs [\ref{KHNT1}] and [\ref{bassett1}], 
and recently re-examined by
Finelli and Brandenberger\cite{FB}.

The gravitational potential
$\Phi$ can be calculated by solving the linearized Einstein equation, 
however, in the case of the adiabatic perturbation with 
scales 
far outside the Hubble radius (the wavenumber $k\rightarrow 0$),
it is convenient to work with
Bardeen parameter, 
\begin{equation} \label{zeta}
\zeta \, \equiv
{2 \over 3}{{\Phi + H^{-1}{\dot \Phi}} \over {1 + w}} + \Phi \, .
\end{equation}
In Eq.(\ref{zeta}) the dot denotes the derivative with respective to 
time, $H$ is the Hubble expansion rate and
$w = p / \rho$ is the ratio of the
pressure to the density of the background.
In the limit of
$k\rightarrow 0$, the Bardeen parameter satisfies\cite{zetaref,MFB}
 \begin{equation}\label{zeta1}
\frac{3}{2}(1+w)H\dot \zeta=0~.
\end{equation}
During the stage of reheating, 
Eq. (\ref{zeta1}) becomes\cite{MFB}:
\begin{equation}\label{dzeta}
  {\dot \phi}^2{\dot \zeta} = 0 ~.
\end{equation}
Recently Finelli and Brandenberger\cite{FB} pointed out that
when the inflaton field oscilates, $\dot \phi=0$ occurs periodically, 
so it is possible to have $\dot \zeta \not= 0$. If it happens, the 
cosmological perturbation will undergo parametric amplification without 
violating causality. 
Specifically, 
they have considered a inflaton model 
with potential $V(\phi) = m^2 \phi^2 / 2$ and solved it 
numerically. 
They found that  
$\zeta$ is constant in time\cite{FB}.

In this paper, we extend their work and consider a general single-field
inflaton potential. We
examine the evolution of $\zeta$ during the
reheating stage, and 
find no changes of $\zeta$ in time. To begin with, we derive the equation
of motion of the Bardeen parameter $\zeta$ in perturbation theory, then we
will take two specific models by numerical calculations to illustrate
our general analytical result.

\section{Perturbation theory and analytic argument}

Working with the longitudinal gauge, 
the perturbed metric can be expressed in terms of the 
gravitational potential $\Phi$
\begin{equation} \label{metric}
ds^2 \, = \, (1 + 2 \Phi)dt^2-a^2(t)(1 - 2 \Phi)dx_idx^i
\, ,
\end{equation}
where $a(t)$ is the scale factor.
The perturbed Einstein equation gives\cite{FB},
\begin{eqnarray}
&&{\ddot \Phi} + 3H{\dot \Phi} + 
\bigl[{{k^2} \over {a^2}} + 2({\dot H} + H^2)\bigr]
\Phi \, = \, \kappa^2 ({\ddot \phi} + H{\dot \phi})\delta\phi~, \label{e2}\\
&&{\ddot{\delta\phi}} + 3H{\dot{\delta\phi}} +
({{k^2} \over {a^2}} + V'')\delta\phi
\,= \, 4{\dot \Phi}{\dot \phi} - 2V'\Phi~, \label{e3}\\
&&{\dot \Phi} + H\Phi \, = \, {1 \over 2}\kappa^2{\dot \phi}{\delta\phi} \, , \label{e4}
\end{eqnarray}
where $\kappa^2 = 8{\pi}G$, $V$ is inflaton potential, 
$\delta \phi$ is the perturbation to the 
field $\phi$, and a prime denotes the derivative 
with respect to $\phi$. 

Inserting (\ref{e4})
into (\ref{e2}) one can obtain the equation of motion for $\Phi$,
\begin{equation} \label{e5}
{\ddot \Phi} + (H - 2{{\ddot \phi} \over {\dot \phi}}){\dot \Phi} +
({{k^2} \over {a^2}} + 2{\dot H} - 
2H{{\ddot \phi} \over {\dot \phi}})\Phi
\, =0~.
\end{equation}

To eliminate the singularities in the equation above when
the inflaton field $\phi$ oscillates, one can make use of
Sasaki-Mukhanov variable
%\cite{MS}
\begin{equation} \label{q}
Q \, \equiv \, \delta\phi + {{\dot \phi} \over H}\Phi~,
\end{equation}
%in term of which,
then Eq.(\ref{e5}) can be re-written as
\cite{KHNT}
\begin{equation}\label{qen}
 \ddot{Q}+3H\dot{Q}+\left[V^{\prime\prime}
   +\frac{k^2}{a^2}+2\left(\frac{\dot H}{H}+3H\right)^{.}   
   \right]Q=0~.
\end{equation}

The Bardeen parameter $\zeta$ is related to $Q$ by
\cite{FB}
\begin{equation}\label{z1}
  \zeta=\frac{H}{\dot \phi}Q~.
\end{equation}

In the expanding universe, the inflaton field
$\phi$ satisfies the equation of motion,
\begin{equation}\label{back}
  \ddot {\phi} + 3H\dot{\phi} +V^{\prime} =0~.
\end{equation}
Differentiating (\ref{back}) with respect to $\alpha\equiv \ln a$
(note that $\dot{\alpha}=H$),
we get
\begin{equation}\label{back1}
  H^{-1}({\phi})^{\cdot\cdot\cdot} 
  +3\ddot{\phi}+(V^{\prime\prime}+3\dot{H})
  H^{-1}\dot{\phi}=0~,
\end{equation}
where $(~)^{\cdot\cdot\cdot}\equiv \frac{d^3}{dt^3}$.

Since $\dot{H}=-\kappa^2\dot{\phi}^2/2$\cite{MFB},
and $\ddot{H}=-\kappa^2\dot{\phi}\ddot{\phi}~$,  
we have the following
relation 
\begin{equation}\label{back3}
  2H^{-2}\dot{H}\ddot{\phi}-H^{-2}\ddot{H}\dot{\phi}=0~.
\end{equation}

Substracting (\ref{back3}) from (\ref{back1}), 
and simplifying it, 
we can obtain 
\begin{equation}\label{qben1}
  \left(\frac{\dot\phi}{H}\right)^{\cdot\cdot}
  +3H\left(\frac{\dot\phi}{H}\right)^{\cdot}+\left[V^{\prime\prime}
   +2\left(\frac{\dot H}{H}+3H\right)^{\cdot}
    \right]\frac{\dot\phi}{H}=0~.
\end{equation}    

Differentiating $Q=\left(\frac{\dot{\phi}}{H}\right)\zeta$ 
with respect to time $t$, we have
\begin{eqnarray}
&&\dot{Q}=\left(\frac{\dot{\phi}}{H}\right)^{\cdot}\zeta+
\left(\frac{\dot{\phi}}{H}\right)\dot{\zeta}~,\\
&&\ddot{Q}=\left(\frac{\dot{\phi}}{H}\right)^{\cdot\cdot}\zeta+
2\left(\frac{\dot{\phi}}{H}\right)^{\cdot}\dot{\zeta}
+\left(\frac{\dot{\phi}}{H}\right)\ddot{\zeta}~.
\end{eqnarray}

Plugging $Q$, $\dot{Q}$ and $\ddot{Q}$ above  
into Eq.(\ref{qen}),
and making use of Eq.(\ref{qben1}), finally we 
obtain an equation of motion for 
$\zeta$ 
\begin{equation}\label{f1}
  \frac{\dot \phi}{H}\ddot {\zeta}+\left(2\frac{\ddot{\phi}}{H}-
2\frac{\dot H}{H^2}\dot {\phi}+3\dot{\phi}\right)
\dot{\zeta}+ \frac{k^2}{a^2}\left(\frac{\dot \phi}{H}\right)\zeta=0~.
\end{equation}

Clearly, the solution of Eq.(\ref{f1}) is that  
$\dot {\zeta} = 0$ when $\dot \phi=0$
(note that $\ddot \phi\ne 0$ at the time when $\dot \phi=0$).
On the other hand, for $\dot {\phi}=0$, Eq.(\ref{dzeta}) gives rise to $\dot{\zeta}=0$.
We conclude that $\zeta$ keeps unchanged during reheating. We should point out that
the potential $V$ in our analytical proof above is not specified, 
and our result applies for general single-field inflaton models. 

\section{Numerical examples}

To illustrate the analytical result in the last section,
we take two specific models as examples by directly solving 
the perturbed Einstein equation with numerical calculations. 
The first model is $V(\phi)=\lambda \phi^4/4$, the second
one is a massive inflaton with 
self-coupled interaction\footnote{The model
considered by Finelli and Brandenberger [\ref{FB1}] corresponds to the
limit of $\lambda \rightarrow 0$. },
$V(\phi)=m^2 \phi^2 / 2 + \lambda \phi^4 / 4$~.
In the latter model two parameters are 
introduced. One is the inflaton mass $m$, 
another is the self-interaction
coupling constant $\lambda$. 
In our numerical calculations, for this model we 
take $m^{-1}$ to be the units of 
the time and leave $\lambda / m^2$ as a free parameter,
at the same time, we only choose $\lambda / m^2 = 1 \times 10^{-3}m_{pl}^{-2} ~, 1m_{pl}^{-2} ~, 1 \times 10^3m_{pl}^{-2}$ 
( where $m_{pl}$ is the Planck mass)
as illustrations.

Fig.1 and Fig.2 present the
evolution of $Q$ and $\zeta$ for these two models,
from which we can see that $Q$ does not change over a period of time and 
$\zeta$ does not change during the zero crossing of 
$\dot \phi$ in the reheating stage (Note that Eqs.(\ref{zeta1}) and (\ref{dzeta}) hold only for the adiabatic perturbation with the wavenumber $k\rightarrow 0$, and we take $k = 0$ for simplicity in the 
numerical calculations.).

\section{Discussion}

In this paper we have studied the evolution of
perturbations in the inflationary cosmology and found no additional growth of gravitational
fluctuations due to the oscillating inflaton field during reheating. Our
result is valid for any single-field inflaton potential. 
For the multiple-field models 
Bassett et al. \cite{BGMK} recently pointed
out that there are possibilities of amplification 
of long wavelength perturbation. This important open issue deserves further study.
~~\\
~~\\

We thank Robert Brandenberger for discussions. This work was supported in 
part by the Natural National Science Fundation of China.

{}

\newpage
\vskip 0.4cm
{\bf Figure Captions}

~~\\
Fig.1
Evolution of $Q$ and $\zeta$ as a function of time in the model
$V(\phi) = \lambda \phi^4 / 4$.
The initial condition is chosen as $Q=-1$ when $\phi=0.2m_{pl}$.
The solid and dashed lines represent
the evolution of $Q$ and $\zeta$ respectively.
Time is expressed in units of
${ ( m_{pl}{\sqrt{\lambda}} )}^{-1}$.\\

~~\\
Fig.2
Evolution of $Q$ and $\zeta$ as a function of time
for a massive
inflaton
$V(\phi) =m^2\phi^2/2+ \lambda \phi^4 / 4$.
(a), (b), and (c) correspond to
$\lambda/m^2=1\times 10^{-3}m_{pl}^{-2},~ 1m_{pl}^{-2},~ 1\times 10^{3}m_{pl}^{-2}$ respectively. Time is in units of $m^{-1}$.

\newpage
\begin{figure}[thb]
\epsfysize=4in
\epsffile{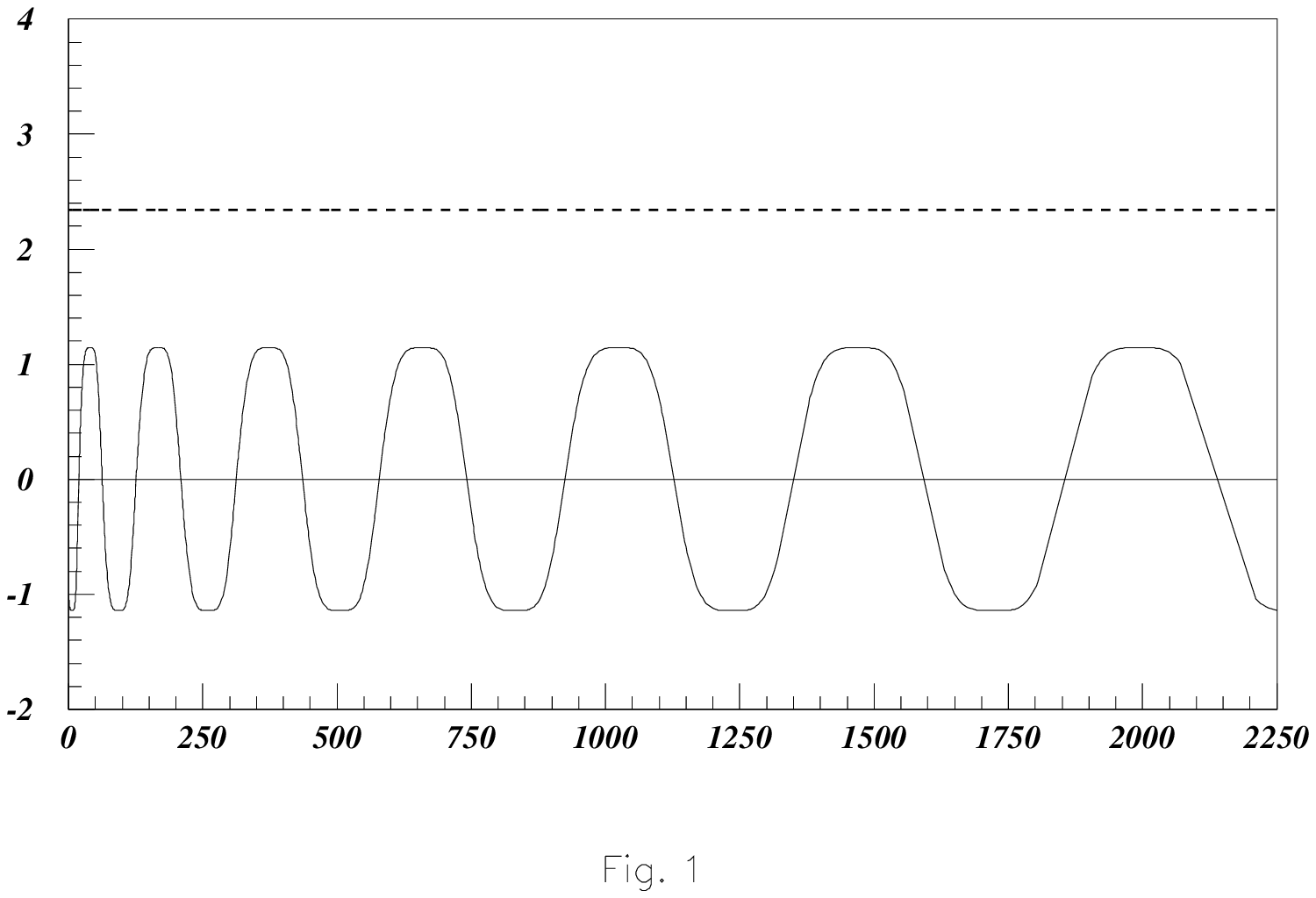}
%\caption[]{
%}
\end{figure}
\newpage
\begin{figure}[thb]
\epsfysize=4in
\epsffile{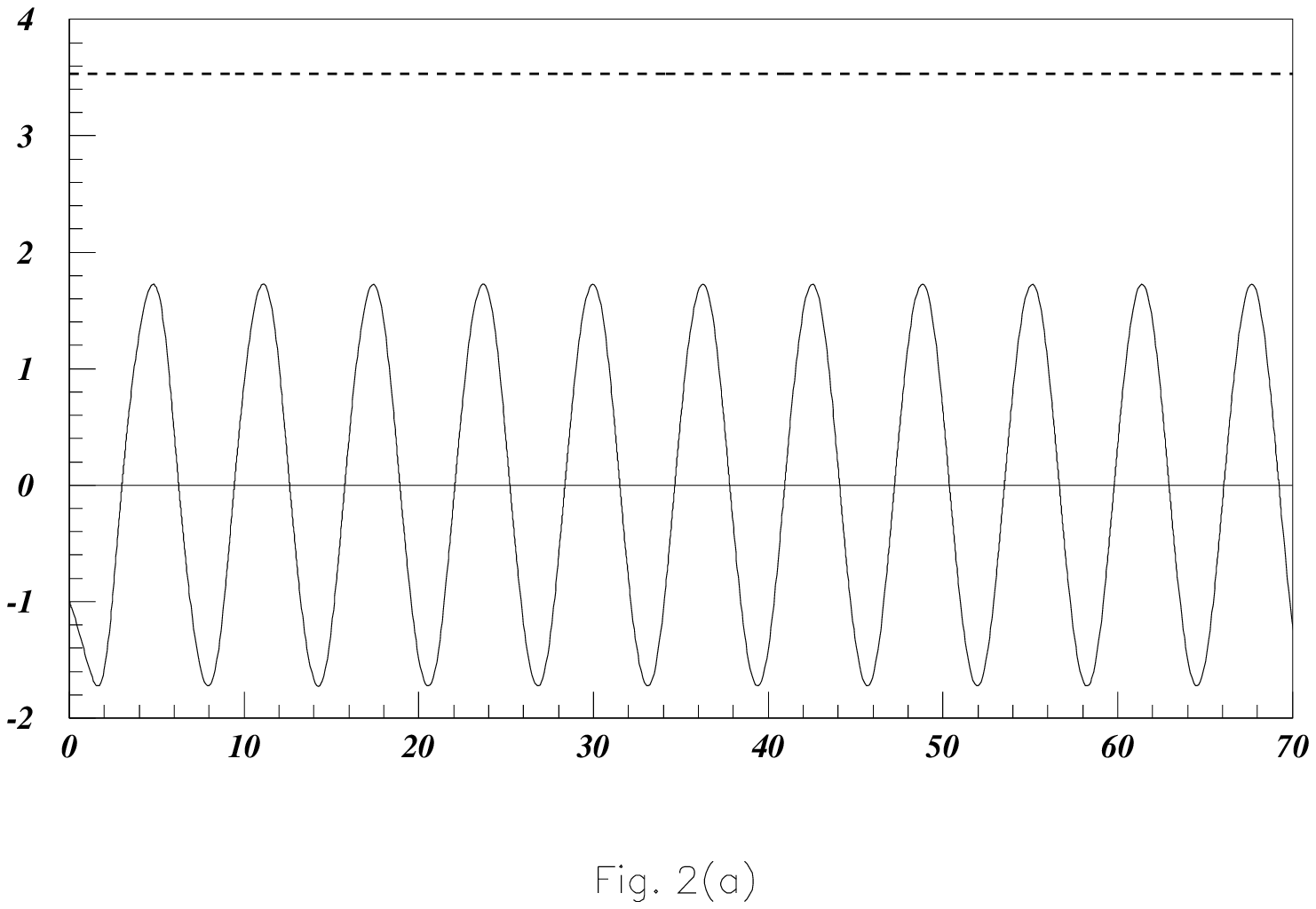}
%\caption[]{
%}
\end{figure}
\newpage
\begin{figure}[thb]
\epsfysize=4in
\epsffile{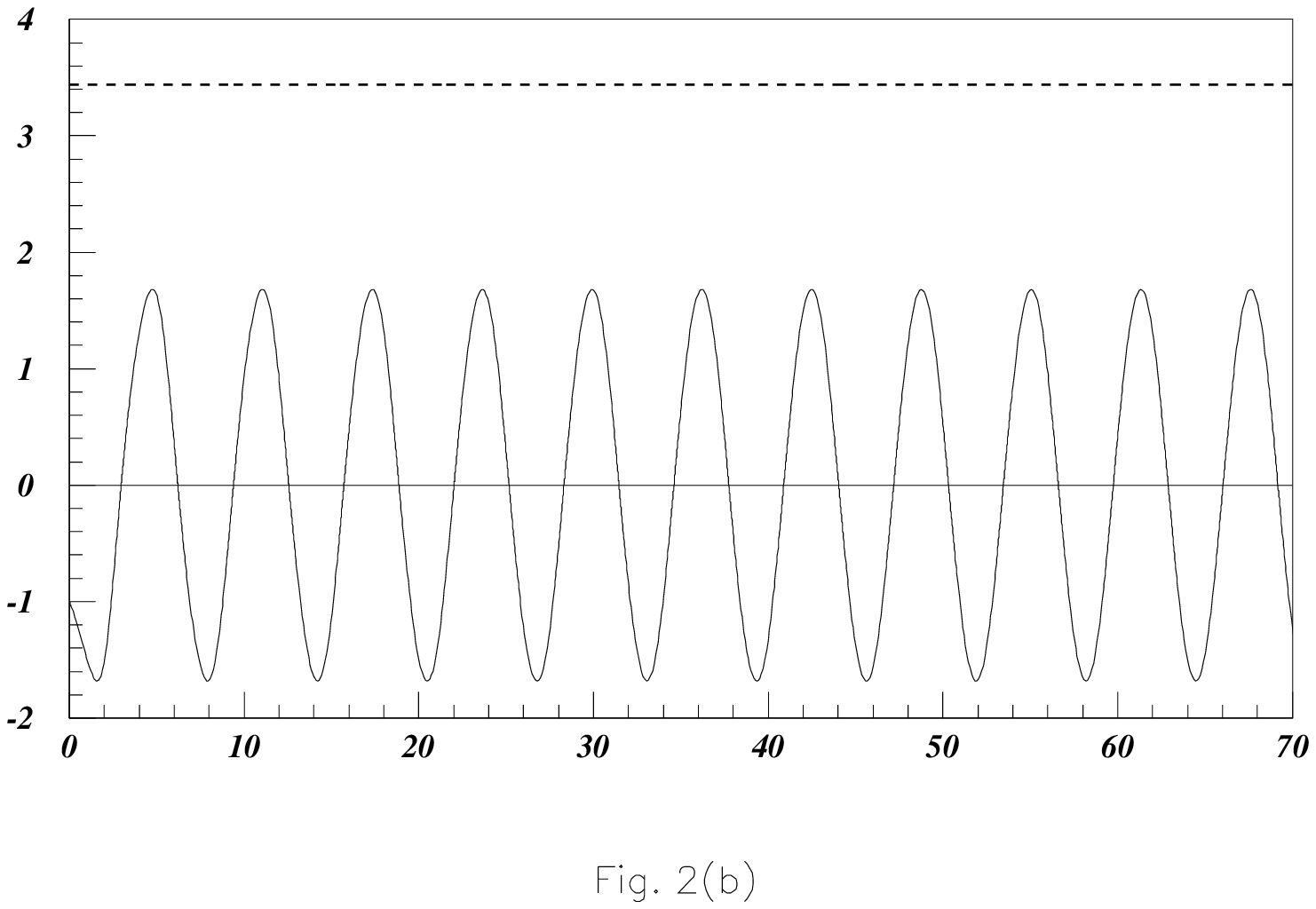}
%\caption[]{
%}
\end{figure}
\newpage
\begin{figure}[thb]
\epsfysize=4in
\epsffile{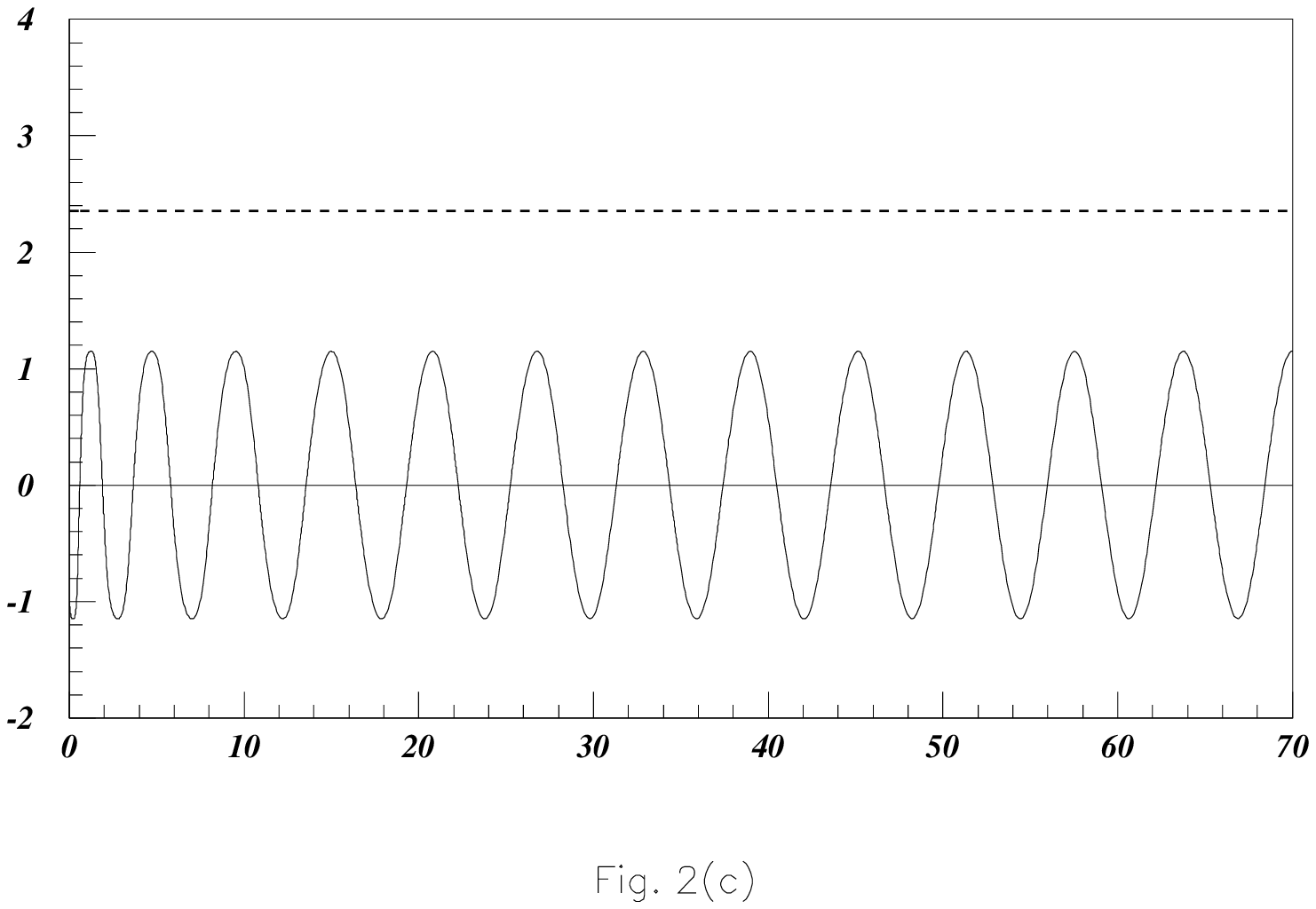}
%\caption[]{
%}
\end{figure}
~~\\
~~\\
\end{document}